\documentstyle[11pt,amssymb]{article}

\makeatletter
\@addtoreset{equation}{section}
\makeatother

\def\dalemb#1#2{{\vbox{\hrule height .#2pt
        \hbox{\vrule width.#2pt height#1pt \kern#1pt
                \vrule width.#2pt}
        \hrule height.#2pt}}}
\def\square{\mathord{\dalemb{6.8}{7}\hbox{\hskip1pt}}}

\def\cA{{\cal A}}
\def\B{{\cal B}}

\def\0{{\sst{(0)}}}
\def\1{{\sst{(1)}}}
\def\2{{\sst{(2)}}}
\def\3{{\sst{(3)}}}
\def\4{{\sst{(4)}}}
\def\5{{\sst{(5)}}}
\def\6{{\sst{(6)}}}
\def\7{{\sst{(7)}}}
\def\8{{\sst{(8)}}}

\def\G{{\cal G}}

\def\ep{\epsilon}
\def\td{\tilde}
\def\wtd{\widetilde}

\let\a=\alpha \let\b=\beta \let\g=\gamma  
    
     \let\r=\rho

\def\nn{\nonumber} \def\bd{\begin{document}} \def\ed{\end{document}}
\def\ds{\documentstyle} \let\fr=\frac \let\bl=\bigl \let\br=\bigr
\let\Br=\Bigr \let\Bl=\Bigl 
\let\bm=\bibitem
\let\na=\nabla
\let\pa=\partial \let\ov=\overline 
\newcommand{\be}{\begin{equation}} 
\newcommand{\ee}{\end{equation}} 
\def\ba{\begin{array}}
\def\ea{\end{array}}
\def\ft#1#2{{\textstyle{{\scriptstyle #1}\over {\scriptstyle #2}}}}
\def\fft#1#2{{#1 \over #2}}
\def\del{\partial}
\def\sst#1{{\scriptscriptstyle #1}}
\def\oneone{\rlap 1\mkern4mu{\rm l}}
\def\ie{{\it i.e.\ }}
\def\etc{{\it etc.\ }}
\def\via{{\it via}}
\def\semi{{\ltimes}}
\def\str{{\rm str}}
\def\jm{{\rm j}}
\def\im{{\rm i}}
\def\mapright#1{\smash{\mathop{-\!\!\!-\!\!\!-\!\!\!-\!\!\!-\!\!\!
             \longrightarrow}\limits^{#1}}}
\def\maprightt#1#2{\smash{\mathop{-\!\!\!-\!\!\!-\!\!\!-\!\!\!-\!\!\!
             \longrightarrow}\limits^{#1}_{#2}}}

\def\cramp{\medmuskip = 2mu plus 1mu minus 2mu}
\def\cramper{\medmuskip = 2mu plus 1mu minus 2mu}
\def\crampest{\medmuskip = 1mu plus 1mu minus 1mu}
\def\uncramp{\medmuskip = 4mu plus 2mu minus 4mu}

\newcommand{\ho}[1]{$\, ^{#1}$}
\newcommand{\hoch}[1]{$\, ^{#1}$}
\newcommand{\bea}{\begin{eqnarray}} 
\newcommand{\eea}{\end{eqnarray}} 
\newcommand{\ra}{\rightarrow}
\newcommand{\lra}{\longrightarrow}
\newcommand{\Lra}{\Leftrightarrow}
\newcommand{\ap}{\alpha^\prime}
\newcommand{\bp}{\tilde \beta^\prime}
\newcommand{\tr}{{\rm tr} }
\newcommand{\Tr}{{\rm Tr} } 
\newcommand{\NP}{Nucl. Phys. }
\newcommand{\tamphys}{\it Center for Theoretical Physics\\
Texas A\&M University, College Station, Texas 77843}
\newcommand{\ens}{\it Laboratoire de Physique Th\'eorique de l'\'Ecole
Normale Sup\'erieure\hoch{2,3}\\
24 Rue Lhomond - 75231 Paris CEDEX 05}
\newcommand{\upenn}{\it Department of Physics and Astronomy\\
University of Pennsylvania, Philadelphia, Pennsylvania 19104}

\newcommand{\auth}{E. Lima\hoch{\dagger}, H. L\"u\hoch{\dagger1},
B.A. Ovrut\hoch{\dagger2} and C.N. Pope\hoch{\ddagger3}}

\begin{document}
\begin{flushright}
\hfill{CTP TAMU-39/99}\\
\hfill{UPR-861-T}\\
\hfill{hep-th/9909184}\\
\hfill{September 1999}\\
\end{flushright}

%\vspace{15pt}

\begin{center}
{ \large {\bf Supercharges, Killing Spinors and 
                Intersecting Gauge Five-branes }}

\vspace{10pt}
\auth

\vspace{10pt}

{\hoch{\dagger}\upenn}

\vspace{10pt}
{\hoch{\ddagger}\tamphys}

\vspace{40pt}

\underline{ABSTRACT}
\end{center}

     We obtain new solutions where a string and a pp-wave lie in the
common worldvolume directions of the non-standard intersection of two
gauge 5-branes in the heterotic string.  The two 5-branes are
supported by independent $SU(2)$ Yang-Mills instantons in their
respective (non-overlapping) transverse spaces.  We present a detailed
study of the unbroken supersymmetry, focusing especially on a
comparison between a direct construction of Killing spinors and a
counting of zero eigenvalues in the annticommutator of supercharges.
The results are in agreement with some previous arguments, to the
effect that additional zero eigenvalues resulting from a
``fine-tuning'' between positive-energy and negative-energy
contributions from different components in an intersection are
spurious, and should not be taken as an indication of supersymmetry
enhancements.  These observations have a general applicability that
goes beyond the specific example we study in this paper.

{\vfill\leftline{}\vfill
%\vskip 5pt
\footnoterule
{\footnotesize \hoch{1} Research supported in part by DOE grant 
DE-FG02-95ER40893 \vskip -12pt} \vskip 14pt
{\footnotesize \hoch{2} Research supported in part by DOE grant
DE-AC02-76ER03071 \vskip -12pt} \vskip 14pt
{\footnotesize  \hoch{3} Research supported in part by DOE 
Grant DE-FG03-95ER40917.\vskip  -12pt}}

\pagebreak
\setcounter{page}{1}

\section{Introduction}

   Over the years, a large menagerie of BPS $p$-brane solutions of the
various supergravities has been discovered.  These include the basic
half-supersymmetric solutions such as the string \cite{dghr} and
5-brane \cite{strom} in $D=10$, and the membrane \cite{dust} and
5-brane \cite{guven} in $D=11$.  In addition, there are BPS solutions
that preserve smaller fractions of supersymmetry, which admit an
interpretation as intersections \cite{tsey} of the basic
half-supersymmetric building blocks.

    The intersections themselves can be divided into two broad
categories.  Firstly, there are the ``standard'' intersections, which
can all be interpreted, by means of a toroidal reduction to some
sufficiently low dimension, as multi-charge $p$-branes for some single
specific $p$, where the charges are carried by different field
strengths of the lower-dimensional theory.  The $p$-brane for each
charge species by itself oxidises back to one specific component of
the intersection in the higher dimension, with the full set of
intersecting objects arising when all the different charge species are
turned on.  This draws attention to the fact that one should also
include BPS configurations that go somewhat beyond what one would
normally call a ``$p$-brane.''  If the charge-carrying field in the
lower dimension is a Kaluza-Klein vector, then back in the original
higher dimension it will have become part of the off-diagonal
structure of the internal part of the metric.  If the Kaluza-Klein
vector carried an electric charge in the lower dimension, this will
give rise to a higher-dimensional metric with a pp-wave propagating
along one of the internal directions.  If, on the other hand, the
Kaluza-Klein field carried a magnetic charge in the lower dimension,
then, from the higher-dimensional standpoint, the metric will have a
Taub-NUT-like monopole structure, sometimes simply called a ``NUT''
for short.  Various $p$-brane examples can be found in \cite{dkl} and
references therein.  Four-dimensional black hole solutions were
classified in \cite{cveticyoum}.  The intersection rules for
$p$-branes in $D=11$ and $D=10$ were classified in \cite{bdejs}.  The
classification of $p$-branes and standard intersections in maximal
supergravities for $2\le D\le 11$ was given in \cite{lpsol,classp}.

     The second category of intersecting solutions consists of what
may be called ``non-standard'' intersections.  These are examples
where there is no simple lower-dimensional interpretation as a
multi-charge $p$-brane.  This is because in these solutions the
harmonic functions for the intersecting ingredients are all
independent of the overall transverse space.  The first such example
was constructed in \cite{khuri}, and further examples were studied in
\cite{bbj,gkt,bdejs,bps}.  A four-dimensional solution with three
perpendicular intersecting membranes (domain walls in $D=4$) was
constructed in \cite{lattice}, admitting an interpretation as a
cosmological lattice universe model.  In fact three is the maximal
number of intersections (with all pair-wise intersections
non-standard) that can occur in supergravity theories.  The
solutions that we shall be constructing in this paper involve a
combination of standard and non-standard intersections.

    Intersections in $D=10$ can arise in both the type II theories and
also in the heterotic theory.  Many of the solutions in the heterotic
theory can also be viewed as solutions in the type II theories, since
the subset of the latter that make use only of the NS-NS fields can be
transferred across directly as solutions in the heterotic string.
However, there are also further possibilities in the heterotic string,
owing to the presence of the Yang-Mills fields.  One possibility is to
find solutions in which the Yang-Mills fields themselves carry charges
that play a r\^ole in supporting the $p$-brane or intersection.
However, in such circumstances one typically finds that the solution
will not be a BPS one.  Another possibility is to use the Yang-Mills
fields to construct an instanton configuration in a four-dimensional
transverse space, which can act as a non-singular source in place of
the more customary point-charge singular sources in the harmonic
functions describing the solution.  One can think of the Yang-Mills
fields here as playing the r\^ole of a ``regulator,'' which smears out
the point-charge singularities.  The first example of such an
instanton-supported soliton was the ``gauge 5-brane'' constructed in
\cite{strom}.  A dyonic string in $D=6$, where both the electric and
magnetic charges are supported by Yang-Mills instantons, was
constructed in \cite{instdyon}.  Such a configuration can also support
a pp-wave propagating in the string world sheet \cite{instdyonwave}.

    The new solutions that we shall construct in this paper involve
many of the ingredients mentioned above.  Specifically, what we obtain
is a ten-dimensional solution of the heterotic theory, describing the
intersection of two 5-branes, a string, and a pp-wave.  If the 5-brane
charges are turned off, the string/wave intersection is of the
``standard'' type, which can be viewed as a 2-charge black hole in
$D=9$, supported by the winding vector and the Kaluza-Klein vector
respectively.  We shall first obtain the more general intersections,
with the two 5-branes, a string and a pp-wave, as solutions using only
the fields of the $N=1$ truncation of type II supergravities; in other
words the metric, the 2-form potential and the dilaton.  As usual in
such solutions, there will be singular sources, corresponding to the
locations of point charges or distributions of charges.  Then, we
shall show that we can generalise the solutions by using the
Yang-Mills fields of the heterotic string to ``smear out'' the two
sets of 5-brane charges.  A novel feature, associated with the fact
that the intersection between the two 5-branes is non-standard, is
that we can introduce separate self-dual Yang-Mills instanton
configurations in the two distinct transverse 4-spaces of the two
5-branes.

    Having obtained the intersecting solutions, we then turn to a
detailed discussion of their supersymmetry.  There has been a rather
confusing literature on the subject of the supersymmetry of
multi-charge $p$-branes, and intersecting $p$-branes, and in the
present paper we attempt to clarify some of these issues.  Although
our discussion will be focused on the particular case of interest
here, it actually provides insights of a more widespread
applicability.  We shall be particularly concerned with addressing the
issue of how one should interpret the occurrence of zero-eigenvalues
of the matrix $\{Q,Q\}$ obtained by anticommuting the supercharges,
and to what extent such zero eigenvalues can be taken as an indication
of the corresponding existence of Killing spinors.  In particular,
when one calculates $\{Q,Q\}$ for BPS configurations involving more
than one kind of charge species, one commonly finds that the number of
zero-eigenvalues can become enhanced for particular fine tunings of
the charges, by cancelling one contribution against
another.\footnote{As opposed to setting charges to zero, which
obviously can enlarge the supersymmetry.}  This would appear to imply
that the supersymmetry can be enhanced at these special charge values,
leading even to preserved supersymmetry fractions such as $\fft34$ or
$\fft78$ in some cases.  Examples of this apparent phenomenon were
found \cite{lpmult} and subsequently laid to rest \cite{taxonomy} in
the past.  In this paper we examine the issue in the context of the
new intersections of two 5-branes, a string and a pp-wave that we
obtain here.  We compare the results from the $\{Q,Q\}$ anticommutator
with the results of direct computation of the Killing spinors, and we
conclude again that the apparent ``supersymmetry enhancements''
suggested by the enlarged numbers of $\{Q,Q\}$ zero-eigenvalues at
special charge values are spurious.  The point is that the derivation
of the connection between zero-eigenvalues of $\{Q,Q\}$ and the
existence of Killing spinors involves certain assumptions about the
global structure of the solutions, including the absence of naked
singularities in the metric, and these assumptions are violated in all
the cases where ``fine-tuning'' of charge parameters enlarges the
number of zero-eigenvalues.  (Some detailed discussion of this point
was given in \cite{dllp}.)  Thus our results here support the previous
contention that no enhancements of supersymmetry occur at fine-tuned
non-vanishing charge values.

   The paper is organised as follows.  In section 2 we construct the
new solutions, comprising a non-standard intersection of two 5-branes
together with a string and a pp-wave, within the framework of the $N=1$
truncation of type II supergravity.  All the ingredients, including
the 5-branes, have singular sources.  Then, in section 3, we
generalise the solutions within the framework of the heterotic theory,
by introducing self-dual Yang-Mills instantons to replace the singular
5-brane sources.  In section 4 we examine the supersymmetry of the
solutions, both from the type IIA or M-theoretic viewpoint and from
the heterotic viewpoint.  We show how the explicit results from
solving the Killing-spinor equations compare with a counting of
zero-eigenvalues in the anticommutator of supercharges, which clarifies
the issue of when the zero-eigenvalue counting procedure gives
trustworthy results for the determination of unbroken supersymmetry.
In section 5, we discuss the near-horizon structure of the
intersecting solutions.  The paper ends with conclusions in section 6.

\section{Intersections with singular sources}

    In this section, we construct the new solution, which is a
non-standard intersection of two 5-branes, a string and a pp-wave,
within the framework of the $N=1$ truncation of the type II strings.
Specifically, it is a solution of the theory described by the
ten-dimensional Lagrangian
%%%%%
\be
{\cal L}_{10} = R {*\oneone} -\ft12{*d\phi}\wedge {d\phi}
 -\ft12 e^{-\phi}\, {*F_\3}\wedge F_\3\,,
\label{d10lag0}
\ee
%%%%%%
and is given by
%%%%%%%
\bea
ds_{10}^2 &=& K^{-3/4}\, H^{-1/4}\, {\wtd H}^{-1/4}\, (- W^{-1}\,dt^2 +
W\, (dx + \eta_w\, (W^{-1} -1)\, dt)^2) \nn \\
&& +K^{1/4}\, H^{3/4}\, {\wtd H}^{-1/4}\, 
(dy_1^2 +\cdots + dy_4^2) \nn \\
&& +K^{1/4}\, H^{-1/4}\, {\wtd H}^{3/4}\, (dz_1^2 +\cdots + dz_4^2)
\ ,\nn \\
\phi&=&-\ft12 \log[K/(H {\wtd H})]\ ,\label{d10sol} \\
F_\3&=&  \eta\, e^{\phi}\, {*({\wtd H}\, dt\wedge dx\wedge 
d^4z\wedge dH^{-1})} 
+ \td\eta\, 
e^{\phi}\, {*(H\, dt\wedge dx\wedge d^4y\wedge d{\wtd H }^{-1})}  \nn \\
&& + \eta_e\, dt\wedge dx\wedge dK^{-1}\ ,\nn
\eea
%%%%%%%
where $H=H(\vec y\,)$, $\wtd H=\wtd H(\vec z\,)$, $K=K(\vec y, \vec
z\,)$ and $W=W(\vec y, \vec z\,)$, and the ten coordinates have been
split as $(t,x,\vec y,\vec z\,)$ with $\vec y=(y_1,y_2,y_3,y_4)$ and
$\vec z=(z_1,z_2,z_3,z_4)$.  The quantities $\eta$\, $\td\eta$,
$\eta_e$ and $\eta_w$ can each independently be chosen to be $\pm1$,
giving a total of 16 equivalent solutions.\footnote{These
16 solutions are all equivalent purely within the framework of the
bosonic sector of the supergravity, but they are not all equivalent when
supersymmetry is taken into account.  This is because field strengths
enter quadratically in the bosonic equations, but linearly in the
supersymmetry transformation rules.}

    We find that equations (\ref{d10sol}) give a
solution provided that the functions $H$ and $\wtd H $ are harmonic
with respect to the flat spaces corresponding to their indicated
coordinate dependences,
%%%%%
\be
\square_{\vec y} \, H=0\,,\qquad \square_{\vec z}\, \wtd H=0\,,
\label{hhhharm}
\ee
%%%%%
while $H_e$ and $W$ satisfy the equations
%%%%%
\be
H^{-1}\, \square_{\vec y} \, K + {\wtd H}^{-1}\, 
 \square_{\vec z}\, K=0\,,\qquad
H^{-1}\, \square_{\vec y} \, W +  
{\wtd H}^{-1}\, \square_{\vec z}\, W=0\,.
\ee
%%%%%
The harmonic functions $H$ and $\wtd H$ are associated with the two
5-branes in the intersection, $K$ is associated with the string, and
$W$ with the pp-wave.  Since there is no overlap between the
coordinate dependences of the $H$ and $\wtd H$ harmonic functions, it
is evident that the intersection is of a non-standard type, since
there is no lower dimension where the configuration could become a
multi-charge $p$-brane.  Furthermore, the functions $K$ and $W$ do not
simply satisfy Laplace equations, but instead satisfy coupled
equations that involve the 5-brane harmonic functions $H$ and $\wtd
H$.  This also is a characteristic feature of non-standard
intersections.  Of course, we can find simple solutions for $K$ and
$W$ by taking
%%%%%
\bea
K(\vec y, \vec z\,) &=& \psi_1(\vec y\,)\, \psi_2(\vec z\,) + 
\psi_3(\vec y\,)
+\psi_4(\vec z\,)\,,\nn\\
W(\vec y, \vec z\,) &=& \chi_1(\vec y\,)\, \chi_2(\vec z\,) + 
\chi_3(\vec y\,)
+\chi_4(\vec z\,)\,,\label{psichi}
\eea
%%%%%
where the functions $\psi_i$ and $\chi_i$ are all harmonic in their
respective subspaces. Note that solutions of this kind, but without
the wave component (\ie with $W=1$), were obtained previously in
\cite{tc}. The solution can also be obtained from the dimensional
reduction of the intersections of a wave, an M2-brane and two 
M5-branes in D=11, which were constructed in \cite{gmt}.

    It is worth remarking that since the Lagrangian (\ref{d10lag0})
can also be viewed as a consistent truncation of the type IIA theory,
we can also regard the above solutions as originating from $D=11$.
In fact we shall exploit this later, when we calculate the
supersymmetry of the solutions.

\section{Instanton-supported intersections}

    We shall now show that we can generalise the above intersection,
by using the Yang-Mills fields of the heterotic string to provide
instanton configurations that will allow the harmonic functions $H$
and $\wtd H$ associated with the 5-branes to be replaced by
non-singular solutions in the heterotic string.  A novel feature that
arises here is that, owing to the non-overlapping nature of the
4-dimensional subspaces coordinatised by $\vec y$ and $\vec z$, where
the functions $H(\vec y\,)$ and $\wtd H(\vec z\,)$ find their support,
we can introduce {\it independent} Yang-Mills instantons for the two
5-branes, using separate $SU(2)$ factors in the $E_8\times E_8$ or
$SO(32)$ gauge group.

    The low-energy effective action of the heterotic string is $N=1$
supergravity in $D=10$, coupled to $E_8\times E_8$ or $SO(32)$
Yang-Mills matter fields.  We shall focus on two orthogonal $SU(2)$
subgroups of $E_8\times E_8$.  The Lagrangian for the bosonic sector
is given by
%%%%%%
\be
{\cal L}_{10} = R {*\oneone} -\ft12{*d\phi}\wedge {d\phi}
 -\ft12 e^{-\phi}\, {*F_\3}\wedge F_\3 - \ft12 e^{-\fft12\phi}\, 
({*G^a_\2} \wedge G^a_\2 + {*\G^{\a}_\2} \wedge \G^{\a}_\2)\ ,
\label{d10lag}
\ee
%%%%%%
where the fields $G^a_\2$ and $\G^{\a}_\2$ are the Yang-Mills field
strengths given by
%%%%
\bea
G_\2^a &=& dB_\1^a + \ft12\epsilon^{abc}\, B_\1^b\wedge B_\1^c\ , \nn \\
\G_\2^\a &=& d\B_\1^\a + \ft12\epsilon^{\a\b\g}\, \B_\1^\b\wedge \B_\1^\g\ ,
\eea
%%%%%%%
and $F_\3$ is the three-form field strength, given by
%%%%%%
\bea
F_\3 &=& dA_\2 + \ft12 B^a_\1 \wedge dB_\1^a +\ft16 \epsilon^{abc}\,
B^a_\1 \wedge B^b_\1 \wedge B^c_\1\ \nn \\
&& + \ft12 \B^\a_\1 \wedge d\B_\1^\a + \ft16 \epsilon^{\a\b\g}\,
\B^\a_\1 \wedge \B^\b_\1 \wedge \B^\g_\1\ .
\eea
%%%%
It satisfies the Bianchi identity
%%%%
\be
dF_\3 = \ft12 G^a_\2 \wedge G^a_\2 + \ft12 \G^\a_\2 \wedge \G^\a_\2\ .
\label{bianchi1}
\ee
%%%%%

    We find that the Lagrangian (\ref{d10lag}) admits solutions of
precisely the same form (\ref{d10sol}) as we obtained in the
Introduction, describing the intersection of two 5-branes, a string
and a pp-wave, except that now the functions $H$, $\wtd H $, $K$ and
$W$ satisfy the more general equations of motion
%%%%%
\bea
&&\square_{\vec y}\, H = -\ft14\, G^a_{ij}\, G^a_{ij}\ ,\qquad
\square_{\vec z}\, {\wtd H} = -\ft14\, \G^\a_{mn}\, \G^\a_{mn}\ ,\nn\\
&&(H^{-1} \square_{\vec y} + {\wtd H }^{-1} \square_z) K = 0\ ,\qquad
(H^{-1} \square_y + {\wtd H}^{-1} \square_z) W = 0\,,
\label{d10eqs}
\eea
%%%%%
where the index contractions in $G^a_{ij}\, G^a_{ij}$ and
$\G^\a_{mn}\, \G^\a_{mn}$ are performed simply using the metrics
$\delta_{ij}$ and $\delta_{mn}$ of the flat four-dimensional
transverse spaces $dy^i\, dy^i$ and $dz^m\, dz^m$ respectively.  The
$SU(2)$ Yang-Mills fields $G_\2^a$ and $\G_\2^\a$ satisfy the
self-duality equations ${*G^a_{ij}}=G^a_{ij}$ and
${*\G^\a_{mn}}=\G^\a_{mn}$ in the four-dimensional flat transverse
spaces respectively, where $*$ denotes Hodge duality in these flat
spaces.  To be precise, we should remark that the bosonic equations
can be satisfied by taking the Yang-Mills fields to be either
self-dual or anti-self-dual (with independent such choices for the two
$SU(2)$ factors).  As with the signs in the expression for the
3-form field strength in (\ref{d10sol}), the different choices that
one makes can impinge upon the supersymmetry of the solutions, as we
shall see later.

          The equations for $K$ and $W$ depend on $H$ and
$\wtd H$.  In this paper, we shall focus on solutions where
$K$ and $W$ are independent of $H$ and $\wtd H$.  Thus, as in
(\ref{psichi}), we may simply take solutions for $K$ and $W$ built
from harmonic functions in the two subspaces:
%%%%%%%%
\bea
K &=&\Big(1 + \sum_\a\fft{2Q_e^\a}{|\vec y-\vec y_\a|^2}\Big)
\Big(1 + \sum_{\a'}\fft{2{Q}_e^{\a'}}{|\vec z-\vec z_{\a'}|^2}\Big)
+\sum_{\a''}\fft{2Q_e^{\a''}}{|\vec y-\vec y_{\a''}|^2} +
\sum_{\a'''}\fft{2 Q_e^{\a'''}}{|\vec y-\vec y_{\a'''}|^2}\,,\nn\\
W &=&\Big(1 + \sum_\b\fft{2P_w^\b}{|\vec y-\vec y_\b|^2}\Big)
\Big(1 + \sum_{\b'}\fft{2 P_w^{\b'}}{|\vec z-\vec z_{\b'}|^2}\Big)
+\sum_{\b''}\fft{2P_w^{\b''}}{|\vec z-\vec z_{\b''}|^2} +
\sum_{\b'''}\fft{2 P_w^{\b'''}}{|\vec z-\vec z_{\b'''}|^2}\,.
\eea
%%%%
Our notation here is that $\vec y_\a$, $\vec y_{\a'}$, \etc, denote
independent sets of instanton locations for each type of index $\a$,
$\a'$, \etc.  Likewise, the quantities $Q_e^\a$, $Q_e^{\a'}$, \etc,
denote independent sets of charges at the different sets of locations.

          The equations for the 5-brane functions $H$ and $\wtd H$
have Yang-Mills source terms.  We shall consider the situation where
the source in each equation is an $SU(2)$ Yang-Mills instanton (using
a different $SU(2)$ subgroup for each equation).  The use of
single-charge and certain classes of multi-charge $SU(2)$ instanton
solutions as sources for the 5-brane were discussed in
\cite{instdyon,instdyonwave}.  We have
%%%%%%
\bea
H &=& 1 + \ft14\square_{\vec y}\, \log\Big(1 +
\sum_{\a=1}^N\fft{\lambda_\a}{|\vec y-\vec y_\a|^2}\Big)
+ \sum_{\a=1}^N\fft1{|\vec y-\vec y_\a|^2}\,,\nn\\
\wtd H&=&1 + \ft14\square_{\vec z}\, \log\Big(1 +
\sum_{\b=1}^{N'}\fft{\td \lambda_\b}{|\vec z-\vec z_\b|^2}\Big)
+ \sum_{\b=1}^{N'}\fft1{|\vec z-\vec z_\b|^2}
\,.\label{instharm}
\eea
%%%%%
As discussed in \cite{instdyonwave}, the final terms in these
expressions for $H$ and $\wtd H$ serve the purpose of subtracting out
singular-source contributions to the functions $H$ and $\wtd H$.
Clearly, one can always add in any harmonic solution of the
homogeneous equations $\square_{\vec y}\, H=0$ and $\square_{\vec z}\,
\wtd H=0$ to the solutions of the inhomogeneous equations given in
(\ref{d10eqs}).  However, we are interested in the case where the
sources are entirely non-singular, coming only from the Yang-Mills
instantons.  It turns out that the terms involving the logarithms in
(\ref{instharm}) actually include singular-source contributions, and
the final terms in the expressions for $H$ and $\wtd H$ are put in
precisely to subtract these out.

    Thus, as discussed in \cite{instdyonwave}, the functions $H$ and
$\wtd H$ given in (\ref{instharm}) satisfy the equations of motion
throughout the space with no singularities.  There are two types of
phase transition that can occur in certain limits of the instanton
moduli, \ie the instanton sizes and their relative locations.  If the
size of an instanton shrinks to zero, the associated $H$ function
becomes harmonic, with a delta-function singularity at the location of
the instanton, implying that a point-like fundamental 5-brane is
created \cite{instdyonwave}.  The second type of phase transition
occurs if two instantons coalesce, leading to the creation of a
point-like fundamental 5-brane \cite{instdyonwave}.  As we shall see
later, the vanishing of the instanton degrees of freedom, and hence the
creation of the fundamental 5-brane, turns the null area of the
horizon to a non-vanishing one.

\section{Supersymmetry}

    To begin, we shall consider the situation where we take the scale
sizes of the Yang-Mills instantons to zero, so that the function $H$
and $\wtd H$ become harmonic, satisfying (\ref{hhhharm}).  Having
studied the supersymmetry in the framework of type IIA supergravity
and M-theory, and then in the heterotic framework, we shall then
consider the situation when the Yang-Mills instantons replace the
singular sources for the 5-branes.  We shall see that the preserved
supersymmetry will be the same whether or not the vanishing-instanton
limit is taken.

\subsection{M-theory perspective}

  When the Yang-Mills instanton scales are both set to zero, the
solution in the heterotic theory can be embedded into type IIA
supergravity, and hence into M-theory.  We shall make use of this in order
to calculate the explicit conditions for the existence of Killing
spinors, by viewing the configuration as a solution in $D=11$.  Then,
we shall compare these explicit results with what one learns by
studying the matrix of anticommutators of supercharges.  This latter
method can be a useful tool for determining the fraction of unbroken
supersymmetry, although as we shall see, the results that come from it
must be interpreted with care.  Specifically, it can sometimes give a
false indication of ``enhanced'' supersymmetry for special values of
the charges, but these always turn out to be spurious, being
associated with configurations with negative-mass contributions in the
metric and naked singularities.  (Similar issues were discussed
previously in \cite{taxonomy}.)

\subsubsection{Killing spinor construction}

   We may view the configuration comprising the intersection of the
two 5-branes, string  and a pp-wave as a solution of $D=11$
supergravity.  To do this, we oxidise from $D=10$ type IIA to $D=11$
using the standard Kaluza-Klein rules:
%%%%%
\bea
d\hat s_{11}^2 &=& e^{-\fft16\phi} \, ds_{10}^2 + e^{\fft43\phi}\, (d\xi
+ \cA_\1)^2\,,\nn\\
\hat F_\4 &=& F_\4 + F_\3\wedge (d\xi + \cA_\1)\,.
\eea
%%%%%
Thus we find that the solution in $D=11$ is given by
%%%%%
\bea
d\hat s_{11}^2 &=& K^{-\fft23} (H \wtd H)^{-\fft13} (-W^{-1}\, dt^2 +
W\, (dx + (W^{-1}-1)\, dt)^2) \nn\\
&&+ K^{\fft13} H^{\fft23} \wtd H^{-\fft13}\, d\vec y{\,}^2 +
 K^{\fft13} H^{-\fft13} \wtd H^{\fft23}\, d\vec z{\,}^2 +
 K^{-\fft23} (H \wtd H)^{\fft23}\, d\xi^2\,,\nn\\
\hat F_\4 &=& ( {*_y\, dH} +  {*_z\, d\wtd H} + 
       dK^{-1}\wedge dt\wedge dx)\wedge d\xi\,,\label{d11sol}
\eea
%%%%%
where $*_y$ and $*_z$ denote the Hodge duals in the four-dimensional
$y$ and $z$ subspaces, in the metrics $dy^i\, dy^i$ and $dz^m\, dz^m$
respectively.  Note that here the solution describes a non-standard
intersection of two 5-branes, a membrane, and a pp-wave.  We have made
the specific choice of the solution where all four $\eta$ sign
parameters in (\ref{d10sol}) are taken to be $+1$.  We shall discuss
the effects of including the $\eta$ parameters later.

    To solve for the Killing spinors, it is useful first to calculate
the spin connection for the class of metrics
%%%%%
\be
d\hat s_{11}^2 = -e^{2A}\,dt^2 + 
e^{2A}\, W^2\, (dx+ (W^{-1}-1)\, dt)^2 + e^{2B}\, d\vec y{\,}^2 + 
e^{2C}\, d\vec z{\,}^2 + e^{2f}\, d\xi^2\,,
\ee
%%%%%
taking the natural orthonormal basis $e^0 = e^A\, dt$, 
$e^9= W\, e^A\, (dx+(W^{-1}-1)\, dt)$ (where we choose to take 
$x^0=t$ and $x^9=x$); $e^i= e^B\, dy^i$, $e^m= e^C\,
dz^m$; and $e^\# = e^f\, d\xi$. (There should be no confusion between
vielbeins and exponentials! Note that $\#$, pronounced ``ten,'' denotes
the vielbein component in the extra dimension.)  We find that the spin
connection is
%%%%%
\bea
&&\omega_{09} = -\ft12 e^{-B}\, W^{-1}\, \del_i W\, e^i -
\ft12 e^{-C}\, W^{-1}\, \del_m W\, e^m \,,\nn\\
&&
\omega_{0 i} = e^{-B}\, (-\del_i A\, e^0-\ft12 W^{-1}\, \del_i W\, e^9)
\,,\nn\\
&&
\omega_{0 m} = e^{-C}\, (-\del_m A\, e^0-\ft12 W^{-1}\, \del_m W\, e^9)
\,,\label{spincon}\\
&&
\omega_{9i} = e^{-B} (\del_i A + W^{-1}\, \del_i W)\, e^9
    -\ft12  e^{-B}\, W^{-1}\, \del_i W\, e^0\,,\nn\\
&&
\omega_{9m} = e^{-C} (\del_m A + W^{-1}\, \del_m W)\, e^9
    -\ft12  e^{-C}\, W^{-1}\, \del_m W\, e^0\,,\nn\\
&&\omega_{ij} = e^{-B}\, (\del_j B\, e^i - \del_i B\, e^j)\,,\qquad
\omega_{im} = e^{-C}\, \del_m B \, e^i - e^{-B}\, \del_i C\,
e^m\,,\nn\\
&&
\omega_{mn} = e^{-C}\, (\del_n C\, e^m - \del_m C\, e^n)\,,\qquad
\omega_{i\#} = - e^{-B}\, \del_i f\, e^\#\,,\qquad \omega_{m\#} = -
e^{-C}\, \del_m f\, e^\#\,.\nn
\eea
%%%%%

     The supersymmetry transformations in $D=11$ are given by
%%%%%
\be
\delta \psi_A = D_A\, \ep - \ft1{288} \Gamma_A{}^{BCDE}\, \ep\, F_{BCDE}
+ \ft1{36}\, \Gamma^{BCD}\, \ep\, F_{ABCD}\,.
\ee
%%%%%
Consider first the $A=0$ vielbein components of this equation.
Substituting the eleven-dimensional solution (\ref{d11sol}) into this,
and using (\ref{spincon}), we obtain
%%%%%
\bea
\delta\psi_0 &=& K^{\fft13}\, (H \wtd H)^{\fft16}\, 
\del_0\ep \label{delta0}\\
&&- K^{-\fft16} H^{-\fft13}\wtd H^{\fft16} 
\Big[ \ft16 K^{-1} \del_i K (\Gamma_{\mu i} +\ep_\mu{}^\nu \Gamma_{\nu
i\#}) \nn\\
&& \qquad + \ft1{12} H^{-1} \del_i
H (\Gamma_{\mu i} + \ft16 \ep_{ijk\ell} \Gamma_{\mu jk\ell \#})
+\ft14 W^{-1}\, \del_i W\, (\Gamma_{0i} + \Gamma_{9i})\Big]\, \ep \nn\\
&& - K^{-\fft16} H^{\fft16} \wtd H^{-\fft13}
\Big[ \ft16 K^{-1} \del_m K (\Gamma_{\mu m} +
\ep_\mu{}^\nu \Gamma_{\nu m\#}) \nn\\
&& \qquad + \ft1{12} \wtd H^{-1} \del_m
\wtd H (\Gamma_{\mu m} + \ft16 \ep_{mnpq} 
\Gamma_{\mu npq \#}) 
+\ft14 W^{-1}\, \del_m W\, (\Gamma_{0m} + \Gamma_{9m})\Big]\, \ep \,.\nn
\eea
%%%%%
 From this, we see that we shall have solutions of
$\delta\psi_0=0$ if $\del_0 \ep=0$, and the following conditions hold:
%%%%%
\bea
&&(\Gamma_i + \ft16 \ep_{ijk\ell}\, \Gamma_{jk\ell \#})\,
\ep=0\,,\qquad
(\Gamma_m + \ft16 \ep_{mnpq}\, \Gamma_{npq \#})\,
\ep=0\,,\nn\\
&&(\Gamma_0 +  \Gamma_{9\#})\, \ep=0\,,\qquad\qquad\qquad
(\Gamma_0 + \Gamma_9)\, \ep=0\,.\label{killing1}
\eea
%%%%%
Note that the first two conditions come respectively from the
coefficients of $\del_i H$ and $\del_m \wtd H$ in (\ref{delta0}),
while the last two conditions come respectively from the coefficients
of $(\del_i K, \del_m K)$ and $(\del_i W, \del_m W)$.  Thus if any of
the functions $H$, $\wtd H$, $K$ or $W$ is trivial (\ie equal to 1),
then the associated condition in (\ref{killing1}) will be absent.
Note also that in deriving {\it separate} conditions associated with
each function, we have implicitly assumed that the functions are
independent (\ie not proportional to one another).

    Proceeding with the $A=9$, $i$, $m$ and $\xi$ components of
$\delta\psi_A$ in a similar fashion, we find that the conditions for
the existence of Killing spinors are just precisely those already
found in (\ref{killing1}), together with
%%%%%
\be
\ep = K^{-\fft16}\, (H \wtd H)^{-\fft1{12}}\, W^{-\fft14}\,  
\ep_0\,,\label{epsol}
\ee
%%%%%
where $\ep_0$ is a constant spinor.  

    If we take the indices $i$ and $m$ in the $\vec y$ and $\vec z$
spaces to range over the values $(1,2,3,4)$ and $(5,6,7,8)$
respectively, it follows from (\ref{killing1}) that the conditions for
the existence of Killing spinors are that the constant spinor $\ep_0$
must satisfy
%%%%%
\bea
H:&&\ep_0 = -\Gamma_{1234\#}\, \ep_0\,,\nn\\
\wtd H:&& \ep_0 = -\Gamma_{5678\#}\, \ep_0\,,\label{killing3}\\
K:&&\ep_0 = \Gamma_{09\#}\, \ep_0\,,\nn\\
W:&& \ep_0 =  \Gamma_{09}\, \ep_0\,,\nn
\eea
%%%%%
where we have made explicit which condition is associated with which 
metric function.

    Any one of the four equations in (\ref{killing3}) by itself has 16
independent solutions for $\ep_0$.  Thus with just one of the four
charges turned on, the solution preserves $\ft12$ of the
eleven-dimensional supersymmetry.  The number of solutions that one
gets when more charges are turned on depends on various factors,
including one's choice of gamma-matrix conventions.  In particular, one
should bear in mind that the product $\Gamma_{0123456789\#}$ must be
either $+\oneone$ or $-\oneone$.  If we make the convention choice
%%%%%
\be
\Gamma_{0123456789\#} =+\oneone\,,\label{gammaconv}
\ee
%%%%%
then we find, for example, that the conditions from $H$ and from $\wtd
H$ are equivalent, and so introducing the charge for $\wtd H$ as well
as for $H$ would yield no further constraints.  On the other hand, if
the opposite gamma-matrix convention to (\ref{gammaconv}) were chosen,
then introducing $\wtd H$ as well as $H$ would cause all Killing
spinors to be lost.  We shall not enumerate here all the supersymmetry
fractions for the various possible non-vanishing sets of
charges, since the results can be summarised more succinctly later.
Let us just remark for now that with optimally-chosen conventions, one
finds that the solution with all four charges active preserves $\ft18$
of the supersymmetry.

   We observed in section 2 that there are actually 16 independent
solutions to the bosonic equations of motion that follow from
(\ref{d10lag0}), where we allow the independent choice of $+1$ or $-1$
for each of the parameters $\eta$, $\td\eta$, $\eta_e$ and $\eta_w$ in
(\ref{d10sol}).  Thus our specific choice in (\ref{d11sol})
corresponds to $(\eta, \td\eta, \eta_e,\eta_w)=(+1,+1,+1,+1)$.  It is
clear that when we reinstate the parameters, the conditions
(\ref{killing3}) will be replaced by
%%%%%
\bea
H:&&\ep_0 = -\eta\, \Gamma_{1234\#}\, \ep_0\,,\nn\\
\wtd H:&& \ep_0 = -\td\eta\, \Gamma_{5678\#}\, \ep_0\,,\label{killing4}\\
K:&&\ep_0 = \eta_e\, \Gamma_{09\#}\, \ep_0\,,\nn\\
W:&& \ep_0 = \eta_w\, \Gamma_{09}\, \ep_0\,.\nn
\eea
%%%%%
We shall return to a discussion of the possible sign choices later.

    Note that using (\ref{gammaconv}), we can replace the gamma-matrix
combinations $\Gamma_{1234\#}$ and $\Gamma_{5678\#}$ in
(\ref{killing4}) by $\Gamma_{056789}$ and $\Gamma_{012349}$
respectively.  This then means that the indices on the first three
gamma-matrix combinations in (\ref{killing4}) can be viewed as lying
in the world-volumes of the two 5-branes and the membrane
respectively.  The last combination, for $W$, lies in the plane in
which the pp-wave propagates.

\subsubsection{Superalgebra analysis}

     It is now instructive to compare the explicit results that we
have obtained for the Killing spinors with what one learns from the
eleven-dimensional supersymmetry algebra.  We shall follow some of the
notation and conventions of \cite{gaunthull}.   One finds that the
anticommutator of the supercharges $Q$ gives the expression
%%%%%
\be
\{ Q, Q\} = C\, (\Gamma^M\, P_M + \ft1{2!} \Gamma^{M_1 M_2}\, Z_{M_1 M_2}
+ \ft1{5!}\, \Gamma^{M_1\cdots M_5}\, Z_{M_1\cdots M_5})\,,\label{qqorig}
\ee
%%%%%
where $C$ is the charge-conjugation matrix, which can be taken to be
$C=\Gamma^0$, and $Z_{M_1 M_2}$ and $Z_{M_1\cdots M_5}$ are 2-form and
5-form charges.  In the present case, where we have two 5-branes and a
membrane supported by $F_\4$ in $D=11$, the charges will be given by 
the asymptotic
integrals of the three terms in $F_\4$ given in (\ref{d11sol}).  Let
us call the 5-brane and membrane charges $q_5$, $\td q_5$ and $q_2$
respectively. Thus we will have non-zero $Z$'s given by
%%%%%
\be
Z_{12349} = q_5\,,\qquad Z_{56789} = \td q_5\,,\qquad
Z_{9\#} = q_2\,.
\ee
%%%%%
In addition, the pp-wave will contribute to the momentum $P_M$ in its
direction of propagation, and so
%%%%%
\be
P_9 = q_w\,.
\ee
%%%%%

    We may choose a basis for the eleven-dimensional gamma matrices 
where 
%%%%%
\bea
\Gamma_{012349} &=& \hbox{diag}(1,1,1,1,-1,-1,-1,-1)\otimes \oneone_4\,,\nn\\
\Gamma_{056789} &=&
\hbox{diag}(1,1,-1,-1,1,1,-1,-1)\otimes \oneone_4\,,\label{gam2}\\
\Gamma_{09\#} &=&  \hbox{diag}(1,1,-1,-1,-1,-1,1,1)\otimes \oneone_4\,,\nn\\
\Gamma_{09} &=&    \hbox{diag}(1,-1,-1,1,-1,1,1,-1)\otimes \oneone_4\,,\nn
\eea
%%%%%
since $\Gamma_{012349}$, $\Gamma_{056789}$, $\Gamma_{09\#}$ and
$\Gamma_{09}$ all commute with one another.  We therefore find that
%%%%%%%
\crampest
\bea
\{Q,Q\} &=& \hbox{diag}(E-q_5-\td q_5 -q_2 - q_w,
E-q_5-\td q_5 -q_2 + q_w,
E-q_5+ \td q_5 + q_2 + q_w,\nn\\
&&\qquad E-q_5+ \td q_5 + q_2 - q_w,
 E+q_5 -\td q_5 + q_2 + q_w,
  E+q_5 -\td q_5 + q_2 - q_w,\nn\\
&&\qquad E+q_5 +\td q_5 - q_2 - q_w,
E+q_5 +\td q_5 - q_2 - q_w)\otimes \oneone_4\,.\label{qqcom3}
\eea
\uncramp
%%%%%%
where the total energy $E$ is given by
%%%%%
\be
E = p_5 +\td p_5 + p_2 + p_w\,.\label{energy2}
\ee
%%%%%
The quantities $p_5$, $\td p_5$, $p_2$ and $p_w$ are the individual
contributions to the ADM mass coming from the two 5-branes, the
membrane and the pp-wave.  They correspond directly to the overall
asymptotic coefficients of leading-order inverse power-law coordinate
dependences of the four metric functions $H$, $\wtd H$, $K$ and $W$.
The corresponding {\it charges} $q_5$, $\td q_5$, $q_2$ and $q_w$ are
related to them by
%%%%%
\be
(q_5, \td q_5, q_2, q_w) =(\eta\, p_5, \td\eta\, \td p_5, \eta_e\,
p_2, \eta_w\, p_w)\,,\label{pqrel}
\ee
%%%%%
where $(\eta, \td\eta, \eta_e,\eta_w)$ are the 16 $\pm$ sign choices in
the bosonic solutions that we discussed previously.
 
    Consider first the case where  
$(\eta, \td\eta,\eta_e, \eta_w)=(+1,+1,+1, +1)$.  We then find that
the anticommutator (\ref{qqcom3}) is given by
%%%%%
\be
\{ Q, Q\} = 2\, \hbox{diag} (0, p_w,\td p_5+p_2+ p_w, \td p_5 + p_2,  
p_5 + p_2+p_w, p_5+p_2, p_5+\td p_5, p_5+\td p_5+p_w)
\otimes \oneone_4\,.\label{qqcom4}
\ee
%%%%%
Clearly in general, namely with all four parameters non-zero, this
will have just 4 zero eigenvalues, giving a counting of $\ft18$
unbroken supersymmetry that is in precise agreement with our findings
from the explicit solutions for the Killing spinors.  For the various
possible combinations of non-vanishing subsets of parameters $(p_5,
\td p_5, p_2, p_w)$, the corresponding numbers of zero eigenvalues in
(\ref{qqcom4}) can be read off.  In each case it is easy to verify
that the counting agrees precisely with our previous derivation of the
constraint equations (\ref{killing3}) for the Killing spinors,
provided that, as usual, the spurious zero eigenvalues that could
apparently be achieved by allowing negative $p$ parameters are
discarded.

    For a total of 8 of the 16 possible sign choices in (\ref{pqrel}),
the story is similar.  It is best summarised by discussing the
eigenvalues of $\{Q,Q\}$, rather than keeping track of the ordering of
diagonal entries, which are permuted around in the various cases.
Thus we may say that for 8 of the sign choices, we obtain the
eigenvalues
%%%%%
\be
\lambda=
2( 0, p_5+p_2, \td p_5 +p_2, p_5+\td p_5, p_w, p_5+ p_2+p_w, \td p_5+p_2+
p_w, p_5+\td p_5+ p_w)\,,\label{waveevsusy}
\ee
%%%%%
each occurring with degeneracy 4.  (This is the same as the set of
eigenvalues in the specific example (\ref{qqcom4}).)  For these
supersymmetric solutions, the mass/charge relations are given by
%%%%%%
\be
E\equiv p_5+\td p_5 + p_2+p_w= q_2\pm(q_5+\td q_5)\pm q_w\,,\qquad
\hbox{or}\qquad -q_2\pm(q_5-\td q_5)\pm q_w\,.
\ee
%%%%%%%
For the remaining 8 possibilities, the eigenvalues are given by
%%%%%
\be
2( p_2, p_5, \td p_5,p_5+\td p_5 +p_2, p_2 +p_w , p_5+ p_w, \td p_5+
p_w, p_5+\td p_5 + p_2 + p_w)\,,\label{waveevnonsusy}
\ee
%%%%%%
and hence these solutions break all the supersymmetry if all the charges 
are non-vanishing.  For these
non-supersymmetric solutions the mass/charge relations are given by
%%%%%%
\be
E\equiv p_5+\td p_5 + p_2+ p_w= q_2\pm(q_5-\td q_5)\pm q_w\,,\qquad
\hbox{or}\qquad -q_2\pm(q_5+\td q_5)\pm q_w\,.
\ee
%%%%%%%

  Naively, it might seem that we could achieve different, sometimes
unusual, fractions of unbroken supersymmetry by making certain special
non-vanishing choices for the $(p_5, \td p_5, p_2, p_w)$
parameters.  For example, choosing $p_5=\td p_5= -p_2$, with $p_w=0$,
in (\ref{waveevsusy}), we could apparently get 24 zero eigenvalues and
hence $\ft34$ preserved supersymmetry.  On the other hand, we saw no
indication at all from the explicit solutions for the Killing spinors
that such ``supersymmetry enhancements'' could occur at special values
of the charges.\footnote{We {\it did} remark at the stage when we
obtained the conditions (\ref{killing1}) for the existence of Killing
spinors that we were assuming that the functions $H$, $\wtd H$, $K$
and $W$ were not proportional to one another.  It is crucial to
appreciate that the quantities $p_5$, $\td p_5$, $p_2$ and $p_w$ are
precisely the coefficients appearing in these functions, and so there
can be no possibility of two of the functions being proportional in
the kinds of cases we are considering here, where one of their $p$
coefficients is the {\it negative} of the other.}

    The resolution to this puzzle, as discussed in
\cite{taxonomy,dllp}, is that the occurrence of zero eigenvalues in
the anticommutator of supercharges can sometimes give a false
impression of the existence of additional Killing spinors, and here we
have encountered precisely such an example.  The reason why the
supercharge argument is unreliable here is that in order to achieve
the extra zeroes in (\ref{qqcom4}) it was necessary to have at least
one of the contributions $(p_5, \td p_5, p_2, p_w)$ that appear in the
energy (\ref{energy2}) be negative.  This means that the associated
harmonic function will have a negative asymptotic coefficient, and
hence it means that there will be a naked singularity in the metric in
some region.  Under such circumstances, the assumed conditions under
which one derives a relation between zeroes of the anticommutator
(\ref{qqorig}) and unbroken supersymmetry generators are violated, and
so one cannot trust the result.  The fact that no additional Killing
spinors actually arise, as seen from our earlier explicit calculation,
shows that this is indeed what has happened in this case.

\subsection{Heterotic theory perspective}

     To study the fractions of preserved supersymmetry in the
heterotic theory, we can use the same Killing-spinor calculations as
we did before, but now we impose the additional ten-dimensional
chirality condition on the Killing spinor,
%%%%%
\be
\ep_0 = \Gamma_\#\, \ep_0\,.\label{chirality}
\ee
%%%%%
(Of course we might instead impose this condition with a minus sign,
depending on our conventions.)  Thus the number of independent
components of unbroken supersymmetry is determined by solving the
equations (\ref{killing3}), together with (\ref{chirality}).  As
usual, if any of the charges in (\ref{killing3}) is zero, then the
associated condition is omitted.  Rather than stating the results for
the various supersymmetry fractions here, it is more convenient first
to give the discussion of the anticommutator of supercharges in this
heterotic case.  As in the previous M-theory discussion, we again find
that the two approaches agree, provided that we discard any apparent
supersymmetry enhancements that would naively appear to occur in the
$\{Q, Q\}$ eigenvalue calculation when there are negative energy
contributions.

    Since we are not for now concerned with the contributions of the
Yang-Mills fields in the anticommutator algebra $\{Q,Q\}$, we can make
use of the same formalism (\ref{qqorig}) as in $D=11$, but with the
additional requirement that we should project all matrices onto the
positive eigenspace of the chirality operator $\Gamma_\#$.  From
(\ref{gam2}), we see that in the gamma matrix conventions we are using
here $\Gamma_\#$ will be given by
%%%%%
\be
\Gamma_\# = \hbox{diag}(1,-1,1,-1,1,-1,1,-1)\otimes \oneone_4\,.
\label{gamma10}
\ee
%%%%%
The projection onto the positive eigenspace of $\Gamma_\#$
therefore amounts to keeping only the 1'st, 3'rd, 5'th and 7'th
entries in (\ref{qqcom3}).  It turns out that for 4 out of the possible
16 sign choices in $(q_5,\td q_5, q_2,
q_w)=(\pm p_5, \pm \td p_5, \pm p_2, \pm p_w)$ we obtain
eigenvalues
%%%%%
\be
\lambda= 2(0, \td p_5 +p_2, p_5+ p_2 + p_w, p_5+\td p_5)\,,\label{lambda1}
\ee
%%%%%
for $\{Q,Q\}$, each with degeneracy 4.  This expression summarises all
the information about the possible supersymmetry fractions that can be
achieved; as usual, all the quantities $(p_5, \td p_5, p_2, p_w)$
should be considered to be non-negative.   The various supersymmetry
fractions implied by taking all possible non-zero subsets of the  
$(p_5, \td p_5, p_2, p_w)$ agree completely with those obtained by
explicitly solving the Killing-spinor equations. The mass/charge
relation for these supersymmetric solutions is given by
%%%%%%
\be
E\equiv p_5+\td p_5 + p_2+p_w= (q_2 +q_w)\pm(q_5+\td q_5)\,,\qquad
\hbox{or}\qquad -(q_2+q_w)\pm(q_5-\td q_5)\,.
\ee

    Note that the other 12 possible sign choices in $(q_5,\td q_5,
q_2, q_w)=(\pm p_5, \pm \td p_5, \pm q_2, \pm q_w)$ divide into three
further sets of 4, with each set leading to eigenvalues as follows:
%%%%%%
\bea
\lambda &=& (p_5+p_2, \td p_5+p_2, p_w, p_5+\td p_5 + p_w)\,,\nn\\
\lambda &=& (p_2, p_5+p_w, \td p_5 + p_w, p_5+\td p_5 + p_2)\,,
\label{lambda234}\\
\lambda &=& (p_5, \td p_5, p_2+ p_w,p_5+\td p_5 + p_2 + p_w)\,.\nn
\eea
%%%%%
Thus with all four charges turned on, none of these other sets gives
rise to any preserved supersymmetry.  Note also that if we make the
opposite sign choice for the chirality projection, then the same sets
of eigenvalues (\ref{lambda1}) and (\ref{lambda234}) occur, but now
with different combinations of sign choice in $(q_5,\td q_5, q_2,
q_w)=(\pm p_5, \pm \td p_5, \pm q_2, \pm q_w)$ being associated with
each set of eigenvalues.  

      The above example provides another manifestation of a
supersymmetry rule obtained in \cite{classp}.  Namely, when a new
intersecting ingredient is introduced in a set of intersections, if it
breaks a further half of the supersymmetry then the structure of the
eigenvalues of $\{Q,Q\}$ is independent of the sign of the new charge.
If, on other hand, the introduction of the new charge does not break
supersymmetry further, then it would break the supersymmetry
completely if it were instead introduced with the opposite sign.  Thus
for $N$ intersecting objects that preserve $1/2^n$ of the
supersymmetry with $n\le N$, $2^n$ out of the $2^{N}$ possible choices
of solutions are supersymmetric while the rest are non-supersymmetric.

    Finally, in our discussion of supersymmetry in the heterotic
framework, we examine the situation where we include the Yang-Mills
fields, and consider the solutions where $H$ and $\wtd H$ are
non-singular functions with Yang-Mills instanton sources, as given in
(\ref{instharm}).  The supersymmetry transformation rule for the
gravitino is unchanged from the one in the $N=1$ truncation of the
type II theory, and so our previous calculation of the supersymmetry
in the gravitino sector goes through unchanged.  (In the derivation of
the Killing spinors, for example in (\ref{delta0}), it was not
important that the functions $H$ and $\wtd H$ be harmonic.)  However,
we now have to consider also the supersymmetry transformations of the
gaugini $\chi^a$ and $\chi^\a$, which are the superpartners of the
$SU(2)$ Yang-Mills fields $B_\1^a$ and ${\cal B}_\1^\a$ respectively.
These take the form
%%%%%
\be
\delta \chi^a = G^a_{ij}\, \Gamma_{ij}\, \ep\,,\qquad
 \delta \chi^\a = {\cal G}^\a_{mn}\, \Gamma_{mn}\, \ep\,.
\ee
%%%%%
Now, if $G_{ij}^a$ is self-dual or anti-self-dual, we shall therefore
find that the Killing spinors $\ep$ should satisfy
%%%%%
\be
(\Gamma_{ij}\pm \ft12\ep_{ijk\ell}\, \Gamma_{k\ell})\, \ep=0
\ee
%%%%%
respectively, with similar conclusions in the $\vec z$ space for
$\delta\chi^\a$.  Equivalently, we can express these conditions as
%%%%%
\be
\ep = \mp \Gamma_{1234}\, \ep\,,\qquad \ep = \mp \Gamma_{5678}\,
\ep\,.\label{ymcon}
\ee
%%%%%
Bearing in mind that we also have the chirality condition
(\ref{chirality}) (or its opposite), we see that the conditions
(\ref{ymcon}) are of the same form as the ones already encountered in
(\ref{killing4}).  Thus the question of whether the solutions are
supersymmetric or not comes down to the issue of achieving a
proper correlation of signs, with the self-duality or
anti-self-duality choice for the Yang-Mills instantons being
correlated with the chirality sign convention for the spinors of the
heterotic theory.  Provided the signs are properly chosen, the
preserved supersymmetry fractions for the Yang-Mills
instanton-supported solutions in the heterotic theory will be the same
as for their corresponding singular-source limits.

\subsection{Further comments}
    
        In the previous discussion we demonstrated that the
correspondence between the counting of zero-eigenvalues in the
anticommutator $\{Q, Q\}$ and the counting of Killing spinors holds if
the energy contribution from each intersecting ingredient is
non-negative.  One cannot trust any additional zero-eigenvalues that
arise by virtue of having any negative-energy contributions from any
of the intersecting ingredients. It should be emphasised, however,
that one does not {\it always} get the wrong conclusion from the
$\{Q,Q\}$ calculation when there are negative-energy contributions.
For example, for a simple extremal $p$-brane solution preserving
$\ft12$ supersymmetry the analogous result is $\{Q, Q\}= 2(0,p)\times
\oneone_{16}$, and this correctly implies that there will continue to
be 16 unbroken components of supersymmetry even if the mass is taken
to be negative.  (The Killing spinor equation continues to admit 16
solutions, even if one sets the mass negative.)  But the $\{ Q, Q\}$
calculation is giving the ``correct conclusion for the wrong reason''
if the mass is negative.  On the other hand, it seems that in all
cases where there appear to be enhanced supersymmetry fractions for
particular tuned sets of charges that involve negative-energy
contributions, the conclusion is always wrong.

        Although Killing spinor solutions still exist even if an
intersecting ingredient contributes a negative energy, the behavior of
the Killing spinor will become singular.  To see this, recall that the
Killing spinor solution (\ref{epsol}) holds regardless of the detailed
structure of the harmonic functions $K$, $H$, $\wtd H$ or $W$.  When
each intersecting ingredient has positive energy, these harmonic
functions take values between 1 and $\infty$, and hence the Killing
spinors are finite.  On the other hand, if any intersecting object
contributes a negative energy, then the associated harmonic function
takes values between 1 and 0, with a naked singularity occurring at the
latter value.  As can be seen from (\ref{epsol}), the Killing spinor
blows up at the naked singularity. 

     In most examples, requiring the regularity of the Killing spinor
leads to a positive energy contribution, which in turn
implies cosmic censorship \cite{klopp}.  However, as observed in
\cite{dllp}, when a 5-brane with negative mass is supported by a
Yang-Mills instanton of sufficient scale size, the previous singularity
in metric can be smeared out by the instanton, and hence the Killing 
spinor is also well-behaved.  In this case, the continued violation of
the relation between the zero-eigenvalue counting in $\{Q,Q\}$  and the
counting of Killing spinors is caused by the fact that the
energy-momentum tensor of the Yang-Mills instanton fields now violates the
positive-energy condition \cite{dllp}.

        For $N$ intersecting objects that preserve $1/2^N$ of the
supersymmetry, the eigenvalues of the anticommutator $\{Q,\}$ have the
form
%%%%%%
\be
\lambda = (E\pm p_1\pm \cdots \pm p_N)=2(0, p_1, p_2, 
\ldots, p_N, p_1+p_2, p_1+p_3,\ldots, p_1+\cdots p_N)\,,
\ee
%%%
where the total energy is $E=\sum_i p_i$.  Thus the anticommutator
$\{Q,Q\}$ would have negative eigenvalues if any of the objects
contributes a negative energy.  On the other hand, if the $N$
intersecting objects preserve $1/2^n$ of the supersymmetry with $n<N$,
then $\{Q,Q\}$ can still remain positive even if some of the
individual energy contributions are negative, since in these cases,
not all the individual energy contributions appear in isolation as
eigenvalues.  For example, the eigenvalues in (\ref{qqcom4}) will
remain non-negative even if we set $q_2+q_5=0$.  The phenomenon
was observed in \cite{cveticyoum1} in the context of 4-charge black
holes in four-dimensional heterotic string theory, where the
eigenvalues of $\{Q,Q\}$ are $2\{0, p_1+p_2, p_3+p_4,
p_1+p_2+p_3+p_4\}$ and the energy is $E=p_1+p_2+p_3+p_4$.  Thus one
can obtain a massless black hole by setting $p_1+p_2=0$ and
$p_3+p_4=0$, without there being any negative eigenvalues.  Another
example in the literature is the dyonic string in $N=1$ supergravity,
for which $\{Q,Q\}$ has eigenvalues $2\{0, p_1+p_2\}$, with energy
$E=p_1+p_2$ \cite{instdyon,dllp}.  It follows that the dyonic string
becomes tensionless when $p_1+p_2=0$.  (Of course as usual, one should
not take the occurrence of extra zero-eigenvalues in these limits as
indicating enhanced supersymmetry.)  It seems that solutions such as
the above massless ones, where no negative eigenvalues occur in $\{Q,
Q\}$, may have a more solid relation to states in the quantum theory
than ones where masslessness is achieved at the price of negative
eigenvalues.

    In our present case in this paper, from an M-theory perspective,
there is no choice of parameters such that the energy $E= p_5 +\wtd
p_5+ p_2 +p_w$ vanishes while the eigenvalues in the associated
anticommutator (\ref{qqcom4}) all remain positive.  On the
other hand, from the heterotic perspective the eigenvalues in the
superalgebra are truncated to (\ref{lambda1}), and hence it is
possible to obtain a massless solution with purely non-negative
eigenvalues in the associated superalgebra.  Of course achieving the
massless solution requires negative-energy contributions from some of
the intersecting ingredients, and hence either the Killing spinor
blows up, or the Yang-Mills energy density becomes negative.  Such a
pathology may imply a phase transition of the type discussed in
\cite{witten2,instdyon}.

\section{Near-horizon structure}

    For a single-center configuration, the solutions to equations 
(\ref{d10eqs}) can be taken to be:
%%%%%%%
\bea
K &=& (1 + \fft{2Q_e}{r^2})(1 + \fft{2Q'_e}{\r^2})\ ,\qquad
W = (1+ \fft{2Q_w}{r^2}) (1 + \fft{2Q'_w}{\rho^2})\ ,\nn \\
H &=& 1 +\fft{2(r^2 + 2a^2)}{(r^2 + a^2)^2} \ ,\qquad
{\wtd H} = 1 +\fft{2(\r^2 +2b^2)}{(\r^2 + b^2)^2} \ ,\label{d10sols}
\eea
%%%%%%%
where $ r^2=y^i y^i$ and $ \rho^2 = z^m z^m$. 
Here $a$ and $b$ are the sizes of the instantons supporting
each of the 5-branes.

        Let us now consider what happens if the two instanton sizes
$a$ and $b$ vanish.  In this situation, the functions $H$ and $\wtd H$
become harmonic functions.  Let us consider the horizon region with
$r\,\rho\rightarrow 0 $ and $r/\rho$ non-vanishingly finite.  In this
region, the additive constants ``1'' in these two functions $H$ and
$\wtd H$ can be dropped.  Also $K\sim 4Q_eQ'_e/(r^2\,\rho^2) $.  In
this region the dilaton scalar $\phi$ becomes constant, and the metric
becomes (for simplicity, we consider $Q_eQ'_e=1$)
%%%%%
\be
ds_{10}^2 = ds_4^2 + 2d\Omega_3^2 + 2d{\wtd\Omega}_3^2\,,
\label{m4s3s3met}
\ee
%%%%%
where $d\Omega_3^2$ and $d{\wtd \Omega}_3^2$ are the metrics for unit
3-spheres, and 
%%%%%%%%%%
\be
ds_4^2 =\ft14 r^2\, \rho^2\, (-W^{-1}\, dt^2 + W (dx + (W^{-1} -1)\,
dt)^2) + \fft{2dr^2}{r^2} + \fft{2d\rho^2}{\rho^2}\ .\label{d4sol}
\ee
%%%%%%%

      The four-dimensional configuration (\ref{d4sol}) is the solution to
the Lagrangian
%%%%%
\be
e^{-1}{\cal L} = R - \ft12(\del\phi)^2 -\ft12 e^{2\phi}\, (\del\chi)^2
+ m^2\, e^{\phi}\,,
\ee
%%%%%%%
which is the scalar Lagrangian of four-dimensional $SU(2)\times SU(2)$
gauged supergravity, constructed in \cite{fs}.

         Making the coordinate transformation $y=\log(r\,\rho)$ and
$z=\log(r/\rho)$, the metric (\ref{d4sol}) becomes
%%%%%%
\bea
&&ds_4^2=\ft{1}{4}e^{2x}\, (-W^{-1}\, dt^2 + W (dx + (W^{-1} -1)\,
dt)^2) + dy^2 + dz^2\,,\nn\\
&&W= (1 + Q_w\, e^{-y-z}) (1+Q'_w\, e^{-y+z})\,.
\eea
%%%%%
When $Q_w=Q_w'=0$, the four-dimensional metric becomes AdS$_3\times
S^1$, as discussed in \cite{tc}.  (Various intersections whose near
horizon structures give rise to an AdS$_3$ spacetime were given
in \cite{bps}.) When the 1's are dropped from $W$,
the above spacetime becomes K$_3\times S^1$, where K$_D$ denotes the
generalised Kaigorodov metric in $D$ dimensions \cite{clp}.  Note that
K$_3$ is locally equivalent to the BTZ black hole constructed in 
\cite{btz}.  The near-horizon structure of this intersection was also
discussed in \cite{gmt}, from an eleben-dimensional point of view.

       The area of the horizon for the metric (\ref{m4s3s3met}), and
hence the entropy, is proportional to $\sqrt{Q_w\, Q'_w}$.  On the
other hand, when the instanton sizes $a$ and $b$ are non-vanishing,
the area of the horizon, and hence the corresponding entropy, would be
zero.  An analogous phase transition occurs also when two instantons
coalesce, which increases the area of the horizon.  The entropy
associated with the non-vanishing area of the horizon can be
understood from the two-dimensional boundary conformal field theory of
the AdS$_3$ spacetime.

         Note that the Lagrangian (\ref{d10lag}) also admits a
different type of four-object intersection, namely a string, 5-brane,
pp-wave and NUT.  This intersection is of the standard type, and gives
rise to a 4-charge black hole \cite{cveticyoum2} in $D=4$.  The
near-horizon structure is BTZ$\times (S^3/Z_n) \times E_4$
\cite{create}. By contrast, the near-horizon structure of the
four-object intersection discussed in this paper is K$_3\times
S^3\times S^3\times S^1$.

\section{Conclusions}

    In this paper, we have constructed extremal solutions in $D=10$,
comprising the non-standard intersection of two 5-branes together with
a string and a pp-wave.  This can arise as a solution in the $N=1$
truncation of the type II theory, with singular sources for all the
ingredients in the intersection.  This configuration can be oxidised
to a solution in $D=11$, where the string now becomes a membrane
living in the common world-volume directions of the 5-branes.  It can
instead be viewed as a solution in the heterotic theory, in which case
it is possible to replace the singular sources for the 5-branes by
self-dual Yang-Mills instantons.  An unusual feature here is that,
owing to the non-standard nature of the 5-brane intersection, in which
they have non-overlapping 4-dimensional transverse spaces, there can
be a separate $SU(2)$ instanton for each 5-brane.  This solution
therefore makes use of an $SU(2)\times SU(2)$ subgroup of the
$E_8\times E_8$ or $SO(32)$ gauge group of the heterotic string.

    We presented a detailed discussion of the supersymmetry of the
intersecting solution.  In particular, we compared the results from an
explicit construction of the Killing spinors with a counting of the
zero eigenvalues of the anticommutator of supercharges, $\{Q,Q\}$.  We
showed that the two are in agreement, provided one discounts as
``spurious'' the additional zero eigenvalues of $\{Q,Q\}$ that can
arise for special ``tuned'' non-vanishing values for certain of the
charges.  We argued that, as discussed in previous examples in the
literature \cite{taxonomy,dllp}, the naive counting of zero
eigenvalues of $\{Q,Q\}$ can give misleading results, if any of the
components in the intersection is giving a negative contribution to
the total energy.  Having exhibited this phenomenon in specific
examples, the implication is that one should always treat apparent
supersymmetry enhancements seen from supercharge anticommutators with
suspicion, unless there is some compelling argument for why they are
not spurious.

\section*{Acknowledgments}

   We are grateful to Mirjam Cveti\v{c} for valuable discussions on 
Killing spinors and the Bogomoln'yi bound in four-dimensional 
black-hole solutions, and to Joachim Rahmfeld for raising again
a question about supersymmetry enhancements.

\end{document}